\documentclass[9pt,conference]{IEEEtran}

\usepackage[preprint]{waspaa25}

\usepackage{bm} 

\usepackage{amsmath,graphicx}
\usepackage{hyperref}
\usepackage{mathtools}
\usepackage{algorithmicx}
\usepackage{algorithm}
\usepackage{algpseudocode}
\usepackage{xcolor}
\usepackage{bm}
\usepackage{amssymb} 
\usepackage{siunitx}
\usepackage{enumitem}
\usepackage{cleveref}
\usepackage{mathtools}
\usepackage{placeins}
\usepackage{booktabs}
\usepackage{microtype}
\usepackage{todonotes}
\usepackage{lipsum}
\usepackage{snaptodo}
\usepackage{optidef}
\hypersetup{
    colorlinks,
    linkcolor={red!90!black},
    citecolor={blue!90!black},
    urlcolor={blue!90!black}
}
\DeclareMathSymbol{\shortminus}{\mathbin}{AMSa}{"39}
\newcommand{\medminus}{\scalebox{0.6}[0.7]{\(-\)}}
\newcommand{\minus}{\mathchoice{-}{-}{\medminus}{\shortminus}}
\newcommand{\Del}{\mathop{}\!\Delta}
\newcommand{\del}{\mathop{}\!\delta}

\DeclarePairedDelimiter\brackets{[}{]}
\DeclarePairedDelimiter\ceil{\lceil}{\rceil}

\DeclareMathOperator{\matrixRank}{rank}
\DeclareMathOperator{\expectedValue}{\mathbb{E}}
\newcommand{\E}[1]{\expectedValue\brackets{#1}}

\renewcommand{\v}[1]{\bm{#1}}  

\newcommand{\ie}{i.e., }

\long\def\hide#1{}

\NewDocumentCommand{\grad}{e{_^}}{%
  \mathop{}\!
  \mathop{}\mspace{-1mu}
  \nabla
  \IfValueT{#1}{_{\!#1}}
  \IfValueT{#2}{^{#2}}
  \mspace{-1mu}
}

\newcommand\scalemath[2]{\scalebox{#1}{\mbox{\ensuremath{\displaystyle #2}}}}
\newcommand\scalemathinline[2]{\scalebox{#1}{\mbox{\ensuremath{\textstyle #2}}}}

\let\ORGhypersetup\hypersetup
\protected\def\hypersetup{\ORGhypersetup}
\makeatletter
\def\th@plain{%
  \thm@notefont{}
  \itshape 
}
\def\th@definition{%
  \thm@notefont{}
  \normalfont 
}
\makeatother

\usepackage[font=footnotesize,position=bottom]{caption}
\usepackage[font=footnotesize,position=bottom]{subcaption}

\def\CyclicSetTheory{\mathcal{A}}
\def\cycidxth{p}
\def\cycSizeth{P}

\def\cycSize{C}
\def\CyclicSet{\mathcal{A}_1} 
\def\cycidx{c}

\def\FourierPair{\overset{\scriptscriptstyle \mathcal{F}}{\longleftrightarrow}}
\def\xmod{X_N^{(\alpha_{\cycidx})}} 

\newcommand{\proc}[1]{\{{#1}\}}  
\newcommand{\fourier}[1]{\tilde{#1}}  

\def\deltaalpha{\del \alpha}
\def\deltaomega{\Del \omega}
\def\alphahat{\hat{\alpha}_1(\ell)}

\def\alphasmooth{\bar{\alpha}_1}
\def\alphatilde{\alphasmooth(\ell)}
\def\alphatildelminusone{\alphasmooth(\ell\minus1)}

\title{Cyclic Multichannel Wiener Filter for Acoustic Beamforming}

\name{Giovanni Bologni$^{\sharp}$, 
Richard Heusdens$^{\sharp\,\flat}$
and Richard C. Hendriks$^{\sharp}$\thanks{This work was supported by NWO, the Dutch Research Council.}}
\address{$\,^{\sharp}$ Delft University of Technology, Delft, the Netherlands\\
$\,^{\flat}$ Netherlands Defence Academy, Den Helder, the Netherlands}

\begin{document}

\maketitle

\begin{abstract}
Acoustic beamforming models typically assume wide-sense stationarity of speech signals within short time frames.
However, voiced speech is better modeled as a cyclostationary (CS) process, a random process whose mean and autocorrelation are $T_1$-periodic, where $\alpha_1=1/T_1$ corresponds to the fundamental frequency of vowels.
Higher harmonic frequencies are found at integer multiples of the fundamental.
This work introduces a cyclic multichannel Wiener filter (cMWF) for speech enhancement derived from a cyclostationary model.
This beamformer exploits spectral correlation across the harmonic frequencies of the signal to further reduce the mean-squared error (MSE) between the target and the processed input.
The proposed cMWF is optimal in the MSE sense and reduces to the MWF when the target is wide-sense stationary.
Experiments on simulated data demonstrate considerable improvements in scale-invariant signal-to-distortion ratio (SI-SDR) on synthetic data but also indicate high sensitivity to the accuracy of the estimated fundamental frequency $\alpha_1$, which limits effectiveness on real data.
\end{abstract}

\section{Introduction}
\label{sec:intro}

A noticeable trait of speech is non-stationarity.
To address non-stationarity, recordings are often divided into short segments, which are then modeled as realizations of wide-sense stationary (WSS) processes in applications such as dereverberation and beamforming \cite{kay_fundamentals_1993,gannot_consolidated_2017,li_alternating_2023,moore_compact_2022}.
However, because of the nearly periodic pressure waves generated by the movement of the vocal folds, voiced speech segments do not behave like WSS processes.
Recently, it has been shown that voiced speech can better be modeled as a (wide-sense) \emph{cyclostationary}  (CS) process \cite{bologni_harmonics_2024}.
We will refer to wide-sense cyclostationary processes simply as cyclostationary (CS) processes in the following.
CS processes describe random signals with first- and second-order moments that vary with frequency $\alpha_1$ \cite{gardner_statistical_1986,gardner_cyclostationarity_1994,gardner_cyclostationarity_2006,feher_short_1995}.
A defining characteristic of CS processes is to exhibit statistical correlation across frequencies, in direct contrast to the WSS assumption, which assumes signals to be uncorrelated over frequency \cite{gardner_cyclic_1993,napolitano_-_2020}.
This distinction is particularly relevant for signals such as voiced speech, where harmonic components at integer multiples of the fundamental frequency $\alpha_1$ occur simultaneously \cite[Ch.~8]{moore_hearing_1995}.
In this context, $\alpha_1$ corresponds directly to the fundamental frequency of vowels.

Adaptive filters typically take advantage of temporal correlations in the signals, whereas spatial filters, known as beamformers, exploit spatial correlations.
This work is grounded in the theory of FREquency-SHifted (FRESH) filtering, which reconstructs CS signals corrupted by noise exploiting the statistical correlation between cyclic frequencies \cite{gardner_cyclic_1993}. 
In other words, FRESH filtering leverages spectral correlations at harmonic frequencies to improve reconstruction accuracy. 
FRESH filtering can be extended to multichannel acoustic recordings to exploit spatial and spectral correlations jointly \cite{zhang_reduced-rank_2006}.

In this paper, we propose a novel beamformer, the cyclic multichannel Wiener filter (cMWF).
Similar to the MWF, our beamformer minimizes the mean-squared error (MSE) between the target and the filter output in the short-time Fourier transform (STFT) domain.
Unlike the MWF, both spatial and spectral correlation of the desired signal are considered when deriving the optimal weights.
The unknown target spectral-spatial covariance matrix is estimated using a generalized eigenvalue decomposition followed by a low-rank approximation.
Our experiments show that the cMWF is particularly advantageous in low-SNR contexts, and it enjoys significant scale-invariant signal-to-distortion ratio (SI-SDR) gains if the signal to reconstruct is indeed cyclostationary at the frequencies of interest.
Moreover, the cMWF reduces to the MWF if the signals under analysis are WSS. 
However, the proposed approach is highly sensitive to errors in the estimation of the fundamental frequency of the target signal, which limits the applicability of the cMWF in speech processing.
A Python implementation of all algorithms is available \cite{github_2025}.

\section{Background}\label{sec:signal_model}
Let us begin by introducing some theory of CS processes. We will denote random variables by capitals and the corresponding realizations by small letters.
Matrices are denoted by bold capitals and vectors by bold small letters.
The tilde denotes frequency domain variables.
A real-valued discrete-time random process $\proc{X(n), n\in \mathbb{Z}}$ is \textit{cyclostationary} (CS) in the wide sense if both its mean and covariance function are periodic with some period $\cycSizeth$:
\begin{equation}
    \mu_X(n) = \mu_X(n + \cycSizeth), \quad
    r_X(n, \tau) = r_X(n + \cycSizeth, \tau),
    ~\forall n, \tau \in \mathbb{Z}.
\end{equation}
As the mean and the covariance of a CS process are periodic in $n$ with period $\cycSizeth$, they accept a Fourier series expansion over a set of \emph{cyclic frequencies}
$
    \scalemath{1.}{\CyclicSetTheory = \{\alpha_{\cycidxth}:2 \pi \cycidxth / \cycSizeth,~\cycidxth=0,\ldots, \cycSizeth-1\}}.
$
\def\funPer{\cycSizeth}
\def\cycCorr{c_X(\alpha_{\cycidxth}, \tau)}
The covariance can thus be expressed as
$\scalemathinline{0.98}{r_X(n, \tau) = \sum_{\alpha_{\cycidxth} \in \CyclicSetTheory}  \cycCorr \exp{(j \alpha_{\cycidxth} n)},}$
where the Fourier coefficients, called \emph{cyclic correlations}, are given by
$\scalemathinline{0.98}{
\cycCorr = \funPer^{-1} \sum_{n=0}^{\funPer-1} r_X(n, \tau) \exp{(-j \alpha_{\cycidxth} n)}.}
$
Now, suppose $\cycCorr$ is absolutely summable w.\,r.\,t.~$\tau$ for all $n$ in $\mathbb{Z}$.
By applying a discrete-time Fourier transform to $\cycCorr$, we get a function $S_X(\alpha_{\cycidxth}, \omega)$ of two frequency variables, a \emph{cyclic} frequency $\alpha_{\cycidxth}$ and a \emph{spectral} frequency $\omega$:
$
    S_X(\alpha_{\cycidxth}, \omega) = \sum_{\tau = -\infty}^{\infty} \cycCorr \exp{(-j \omega \tau)}
$
\cite{gardner_cyclostationarity_1994}.
The quantity $S_X(\alpha_{\cycidxth}, \omega)$ is known as \emph{spectral correlation density}, or cyclic spectrum, as for finite-length processes it is also given by:
\begin{align}\label{eq:spec_corr_freq}
    S_X(\alpha_{\cycidxth}, \omega) = \E{\fourier{X}_N(\omega) \fourier{X}_N^*(\omega - \alpha_{\cycidxth})}, 
\end{align}
where $\fourier{X}_N(\omega) = \sum_{n=0}^{N-1} X(n) \exp{(-j\omega n)}$ is the $N$-point Fourier transform of $\scalemath{0.95}{\proc{X(n)}}$.
The spectral correlation density (SCD) boils down to the conventional power spectral density (PSD) for $\cycidxth = 0$.
A key property of CS processes is to exhibit inter-frequency correlations.
In fact, $\fourier{X}_N(\omega_1)$ is correlated with $\fourier{X}_N(\omega_2)$ for $|\omega_1 - \omega_2| = \alpha_{\cycidxth},~\forall\alpha_{\cycidxth} \in \CyclicSetTheory \setminus \{0\}$.
Intuitively, if we measure $\fourier{x}_N(\omega_1)$, we know something about $\fourier{x}_N(\omega_2)$.
In contrast, the spectral components of WSS processes are asymptotically uncorrelated. For example, for white Gaussian noise, we have that
$S_X(\alpha_{\cycidxth}, \omega) = 0$ for all $\alpha_{\cycidxth} \neq 0$.
Notice that all quantities in this section are defined for a single random process, but generalizing the notions to the cross-statistics between multiple processes is straightforward.
\subsection{Estimation of the cyclic spectrum}\label{ssec:est_spec_corr_acp}
The proposed beamformer, which will be introduced in \cref{sec:prop_algo}, requires knowledge of the cyclic spectrum $S_X(\alpha_{\cycidxth}, \omega)$.
However, the definition of the cyclic spectrum in \cref{eq:spec_corr_freq} involves an ensemble expectation.
To estimate the cyclic spectrum of a CS process, one can use the \emph{time-averaged cyclic periodogram} (ACP) algorithm \cite{gardner_measurement_1986}.
Essentially, the ACP replaces the expectation with a time average and coincides with Welch's PSD estimator for $\cycidxth = 0$ \cite{antoni_cyclic_2007}.
The ACP estimator has the desirable property to produce consistent estimates of the SCD even from a single record or realization of the signal.
Other methods for SCD estimation may offer faster computations if knowledge of the SCD at all spectral and cyclic frequencies is required \cite{roberts_computationally_1991,borghesani_faster_2018,alsalaet_fast_2022}.

Let $\proc{X(n), n \in \mathbb{Z}}$ and $\proc{Y(n), n \in \mathbb{Z}}$ be random processes sampled with sampling frequency $f_s$. 
The processes $\proc{X_N(n)}$ and $\proc{Y_N(n)}$ equal $\proc{X(n)}$ and $\proc{Y(n)}$ for $n \in \{0, \ldots, N-1\}$ and are zero otherwise.
Processing these signals in the STFT domain, where the window length $K$ equals the DFT points and the block shift is $R$, yields a total of $L = \ceil{1 + (N - K) / R}$ frames.
Notice that the spectral resolution is determined by the length $K$ of the DFT analysis window, with $\Del \omega \approx f_s / K~[\si{\hertz}]$.
In contrast, the cyclic frequencies $\alpha_p$ are sampled on a finer grid. Their resolution depends on the total length of the signal, giving $\Del \alpha \approx {f_s}/({L R})~[\si{\hertz}]$ \cite{gardner_measurement_1986}.
The frequency shifted components $\fourier{X}(\omega_k - \alpha_{\cycidxth})$ are not $1/K$ separated.
Instead, the frequency translation at the right-hand side of \cref{eq:spec_corr_freq} is achieved by first modulating in the time domain with cyclic frequency $\alpha_{\cycidxth}$ and then transforming to the frequency domain, which takes advantage of the modulation property of the DFT: 
\begin{align}
\scalemath{1.}{
\fourier{X}(\omega_k - \alpha_{\cycidxth}) \FourierPair X(n) e^{j \alpha_{\cycidxth} n}.
}
\end{align}
The modulated signal in the time domain and its STFT counterpart are given by:
\begin{subequations}
\begin{gather}
    \xmod(n) = X_N(n) e^{j n \alpha_{\cycidxth}}, \\
    \fourier{X}(\omega_k - \alpha_{\cycidxth}, \ell) = \sum_{n=0}^{N-1} \xmod(n + \ell R){w}(n)e^{-j n \omega_k},
\end{gather}
\end{subequations}
where $\ell$ is the time-frame index and $w(n)$ represents a window function of length $N$.
The ACP estimate at cyclic frequency $\alpha_{\cycidxth}$ and spectral frequency $\omega_k$ is then given by:
\begin{align}\label{eq:acp_estimator}
    \hat{S}_{YX}(\alpha_{\cycidxth}, \omega_k) = \frac{1}{L}\sum_{\ell=0}^{L-1} \fourier{Y}(\omega_k, \ell) \fourier{X}^*(\omega_k - \alpha_{\cycidxth}, \ell).
\end{align}
Additional details on implementing the ACP estimator on natural data are discussed in \cref{ssec:rec_avg}.

\subsection{Narrowband beamforming}\label{ssec:nb_beamform}
Let us introduce the signal model and briefly review the beamforming theory.
The general goal of beamforming is to estimate the target signal as a linear combination of the noisy inputs. 
Let $\v{x}(\omega_k) = [\fourier{X}_0(\omega_k)~\ldots~\fourier{X}_{M \minus 1}(\omega_k)]^T \in \mathbb{C}^M$ denote noisy and reverberant measurements from a microphone array with $M$ elements in the STFT domain:
\begin{align}\label{eq:sig_mod_nb}
\v{x}(\omega_k) &= 
\fourier{S}(\omega_k)\,{\v{a}}(\omega_k) + \v{v}(\omega_k) 
= {\v{d}}(\omega_k) + \v{v}(\omega_k),
\end{align}
where 
$
{\v{a}}(\omega_k) = \begin{bmatrix}
    1&
    a_1(\omega_k)~
    \ldots~
    a_{M \minus 1}(\omega_k)
\end{bmatrix}^T
$
is the relative transfer function (RTF) between a reference sensor, the first one in this case, and the remaining sensors,
$\fourier{S}(\omega_k)$ is the target signal at the reference microphone,
and $\v{v}(\omega_k)$ is a noise term.
Following the multiplicative transfer function approximation, the late reverberation component is neglected \cite{braun_evaluation_2018}.
\hide{
A popular approach is the minimum variance distortionless (MVDR) beamformer, which is formulated as the solution to:
\begin{mini}|s|[0]
    {\v{w}(\omega_k)}{\E{\|\v{w}^H(\omega_k) \v{x}(\omega_k)\|_2^2},}
    {}
    {\label{eq:mvdr_nb}}{}
    \addConstraint{\v{w}^H(\omega_k) \v{a}(\omega_k)}{= 1},
\end{mini}
where the goal is to reduce the power of the noisy signal as much as possible while retaining signals whose transfer function is $\v{a}(\omega_k)$.
}
The MWF is a well-known beamformer that minimizes the MSE between the (unknown) target and the input signals \cite{doclo_speech_2005}.
To avoid that the norm of the weights becomes too large in presence of estimation errors, an L2 regularization term can be added to the reconstruction loss, yielding the so-called \emph{robust MWF} \cite{li_robust_2003}:
\begin{mini}|s|[0]
    {\v{w}(\omega_k)}{\E{|\fourier{S}(\omega_k) - \v{w}^H(\omega_k) \v{x}(\omega_k)|^2} + \lambda \|\v{w}\|_2^2 ,}
    {}
    {\label{eq:mwf_nb_robust}}{}
\end{mini}
where $\lambda$ is a hyperparameter that balances the contribution of the two loss terms.
The solution to \cref{eq:mwf_nb_robust} is given by
\begin{align}
    \v{w}_{\text{MWF}}(\omega_k) = (\v{R}_x(\omega_k) + \lambda \v{I})^{-1} \sigma_s^2(\omega_k)\v{a}(\omega_k),
\end{align}
where $\v{R}_x(\omega_k) = \E{\v{x}(\omega_k)\v{x}^H(\omega_k)}$ is the noisy covariance matrix and $\sigma_s^2(\omega_k) = \E{|\fourier{S}(\omega_k)|^2}$ is the variance of the target signal.
A possible choice of $\lambda$ is given by \cite{qiang_wu_blind_1996}:
\begin{align}\label{eq:diag_loading}
\lambda = \min(\lambda_{\text{max}}, \max(\lambda_{\text{min}}, \text{trace}~\hat{\v{R}}_{\v{d}}(\omega_k))),
\end{align}
where $\hat{\v{R}}_{\v{d}}(\omega_k)$ is an estimate of the target spatial covariance matrix, such as $\hat{\v{R}}_{\v{d}}(\omega_k) = \hat{\v{R}}_{\v{x}}(\omega_k) - \hat{\v{R}}_{\v{v}}(\omega_k)$, and $\hat{\v{R}}_{\v{v}}(\omega_k)$ is an estimate of the noise covariance matrix.
The constants $\lambda_{\text{min}}$ and $\lambda_{\text{max}}$ are defined in \cref{sec:experiments}.
Diagonal loading constrains the norm of the weight vector and reduces the sensitivity of the beamformer to errors in the statistics \cite{cox_robust_1987}.
This choice of $\lambda$ gives higher loading when the signal power is higher and smaller loading in noise-dominated segments.

\section{Proposed algorithm}\label{sec:prop_algo}
Our goal is to improve the robust MWF introduced in \cref{ssec:nb_beamform} by exploiting frequency correlations in the target signal.
To this end, we extend the narrowband model in \cref{eq:sig_mod_nb} to form the \emph{multiband} model, which incorporates frequency-shifted versions of the received signal.
The frequency shifts are chosen so that the signal exhibits maximal self-correlation after shifting.
Therefore, we focus on the fundamental frequency $\alpha_1$ and its integer multiples, known in acoustics as \emph{harmonic} frequencies.
The set $\CyclicSet$ of modulation frequencies applied to the signal is defined as
\begin{align}\label{eq:cyc_set_exp}
\CyclicSet = \{\alpha_{\cycidx} : \cycidx \, \alpha_1,~\cycidx = 0, \ldots, \cycSize - 1\},
\end{align}
where the number of modulations $\cycSize$ must be less than or equal to the number of harmonics in the signal, \ie $\cycSize \leq \cycSizeth$.
Next, we compute the STFT of each modulated signal and form a long vector $\v{x}(\CyclicSet, \omega_k) \in \mathbb{C}^{M\cycSize}$, by stacking the non-modulated noisy signal together with all modulated versions:
\begin{align*} 
\scalemath{0.98}{
\v{x}(\CyclicSet, \omega_k)^T = 
\begin{bmatrix} 
\v{x}(\omega_k)^T~ \v{x}(\omega_k - \alpha_1)^T~ \cdots~ \v{x}(\omega_k - \alpha_{\cycSize \minus 1})^T 
\end{bmatrix}. 
}
\end{align*} 
From here on, we write $\v{x}(\CyclicSet, \omega_k) = \v{x}$ to represent multiband signals. 
The modulated reverberant signal vector $\v{d}$ and the modulated noise vector $\v{v}$ are constructed similarly, leading to: 
\begin{align} 
\v{x} = \v{d} + \v{v} \in \mathbb{C}^{M\cycSize}. 
\end{align}
Now, notice that $\v{d}$ can be represented by the matrix-vector multiplication $\v{d} = \v{C}\v{s}$,
where 
$
\v{s} = \begin{bmatrix} 
\fourier{S}(\omega_k)~ 
\allowdisplaybreaks \allowbreak
\ldots~ \allowdisplaybreaks \allowbreak
\fourier{S}(\omega_k - \alpha_{\cycSize \minus 1}) 
\end{bmatrix}^T 
$
is the modulated signal at the reference microphone and $\v{C} \in \mathbb{C}^{M\cycSize \times \cycSize}$ contains a frequency-shifted RTF padded with zeroes for each column. 
For example, for $\cycSize=2$ we have
\begin{align}\label{eq:reverberant_target}
\v{d} = \v{C}\v{s} = 
\begin{bmatrix}
\v{a}(\omega_k) & \v{0}_{M(\cycSize-1)} \\
\v{0}_{M(\cycSize-1)} & \v{a}(\omega_k - \alpha_1)
\end{bmatrix}
\begin{bmatrix}
    \fourier{S}(\omega_k) \\
    \fourier{S}(\omega_k - \alpha_1)
\end{bmatrix},
\end{align}
where $\v{0}_A$ represents a zero vector of size $A$.
Let us also define
\begin{align}
    \v{S}_{\v{x}}(\CyclicSet, \omega_k) = \v{S}_{\v{x}} = \E{\v{x}\v{x}^H} \in \mathbb{C}^{M\cycSize\times M\cycSize}
\end{align}
as the spatial-spectral covariance matrix of the noisy signal.
The spatial-spectral covariance matrices of the reverberant target and the noise are defined similarly and denoted by $\v{S}_{\v{d}}$ and $\v{S}_{\v{v}}$.
Based on the multiband signal model, it is possible to extend the robust MWF beamformer to optimally combine the noisy signals across different microphones and frequency shifts.
The extended design is derived as the minimizer of the cost function below, which shares a similar form with \cref{eq:mwf_nb_robust}:
\begin{align}\label{eq:j_wb_mwf}
\scalemath{0.95}{
J(\CyclicSet, \omega_k) = \E{\|\fourier{S}(\omega_k) - \v{w}^H \v{x}\|_2^2} + \lambda \|\v{w}\|_2^2.
}
\end{align}
We use Wirtinger derivatives to obtain the gradient of \cref{eq:j_wb_mwf} \cite{brandwood_complex_1983}. The solution is obtained by setting the gradient with respect to $\v{w}^*$ to zero:
\begin{align}\label{eq:j_wb_mwf_sol}
    \v{w}_{\text{cMWF}} &= \v{S}_{\lambda}^{-1} \v{s}_{\v{x}\fourier{s}}
    = \v{S}_{\lambda}^{-1} \v{s}_{\v{d}\fourier{s}}
    = \v{S}_{\lambda}^{-1} \v{S}_{\v{d}}\,\v{e}_0, 
\end{align}
where we defined 
$\v{S}_{\lambda} = \v{S}_{\v{x}} + \lambda \v{I}$, 
$\v{s}_{\v{x}\fourier{s}} = \E{\v{x}\fourier{S}^*(\omega_k)}$, 
$\v{s}_{\v{d}\fourier{s}} = \E{\v{d}\fourier{S}^*(\omega_k)}$, 
$\v{e}_0 = [1, 0, \ldots, 0]^T$ and the second equality follows from the assumption that the target and the noise are uncorrelated.

\subsection{Estimating statistics}\label{ssec:est_statistics}
In practice, the harmonic frequencies of the signal and the spectral-spatial covariance matrices are unknown.
The modulation set $\CyclicSet$ is found by first estimating the fundamental frequency $\alpha_1$ using an algorithm based on non-linear least squares (NLS) \cite{nielsen_fast_2017}.
The number of modulations $\cycSize$, which is related to the model order of the harmonic signal, is treated as a hyper-parameter and determined from the experiments.
\Cref{ssec:rec_avg} provides additional details on how to handle the estimation of $\alpha_1$ on non-stationary data.
The elements of $\v{S}_{\v{x}}$ are estimated using the ACP method detailed in \cref{ssec:est_spec_corr_acp} per each spectral frequency $\omega_k$, cyclic frequency $\alpha_c \in \CyclicSet$  and microphone pair.
$\v{S}_{\v{x}}$ is estimated from the noisy measurements, while $\v{S}_{\v{v}}$ is estimated from a noise-only segment. 
In most cases, the noise does not exhibit spectral correlation at the cyclic frequencies $\CyclicSet$ associated with the target, \ie only the null cyclic frequency $\alpha_0=\{0\}$ is shared between the target and the noise.
To enforce this assumption, we adjust the ACP estimate as
$\hat{\v{S}}_{\v{v}} \leftarrow \text{blkdiag}{(\hat{\v{S}}_{\v{v}})},$ where $\text{blkdiag}(\cdot)$ extracts the block diagonal of a matrix, where each square block has size $M$. 
This modification retains only the spatial correlation of the noise while forcing its spectral correlation to 0, thereby reducing the number of unknowns in $\hat{\v{S}}_{\v{v}}$ from $M^2 \cycSize^2$ to $M^2 C$.
The target $\hat{\v{S}}_{\v{d}}$ is then estimated using the generalized eigenvalue decomposition (GEVD) of $(\hat{\v{S}}_{\v{x}}, \hat{\v{S}}_{\v{v}})$, retaining only the $\cycSize$ eigenvectors associated with the $C$ largest eigenvalues, where $\cycSize$ is the maximum possible rank of $\v{S}_{\v{d}}$,
since from the definition of $\v{d}$ in \cref{eq:reverberant_target}, following the lines of \cite[Lemma 1]{bologni_wideband_2025}, we have:
\begin{align*}
\scalemath{1.}{
    \matrixRank{\v{S}_{\v{d}}} = \matrixRank{\v{C} \v{S}_s \v{C}^H} \leq \min{(\matrixRank{\v{C}}, \matrixRank{\v{S}_s})} \leq C.
    }
\end{align*}
The resulting blind cMWF is given by:
\begin{align}\label{eq:cmwf_blind}
    \hat{\v{w}}_{\text{cMWF}} = \hat{\v{S}}_{\lambda}^{-1} \hat{\v{S}}_{\v{d}}^{\text{gevd}}\, \v{e}_0,
\end{align}
where $\hat{\v{S}}_{\lambda} = \hat{\v{S}}_{\v{x}} + \lambda \v{I}$.

\subsection{Recursive averaging}\label{ssec:rec_avg}
The estimation methods in \cref{ssec:est_statistics} are valid as long as the fundamental frequency $\alpha_1$ remains fixed.
However, the estimates $\alphahat$ provided by the NLS algorithm varies from frame to frame.
Direct use of $\alphahat$ in computing the covariance matrices would require recomputing the modulated signals and their covariance matrices at every frame, yielding estimates with high variance.
To mitigate this, define the relative temporal variation in the fundamental frequency at frame $\ell$ as:
\begin{align}
\scalemath{1.}{
    \deltaalpha = |\alphahat - \hat{\alpha}_1(\ell - 1)|/(\hat{\alpha}_1(\ell-1) + \epsilon),
    }
\end{align}
where $\epsilon=\num{e-6}$ avoids divisions by 0 and $\alphahat=0$ if the $\ell$th frame is unvoiced.
Next, introduce the smoothed fundamental frequency estimate, $\alphatilde$, which is used to compute the time-dependent cyclic set $\CyclicSet(\ell)$, the modulated signals, and their statistics:
\begin{align}
\scalemath{1.}{
    \alphatilde = 
    \begin{cases}
    \alphahat & \text{if }  D_0 \leq \deltaalpha < D_1, \\
    \alphatildelminusone & \text{otherwise},
    \end{cases}
    }
\end{align}
where $D_0 < D_1$ are real-valued thresholds, and $\alphasmooth(0) = 0$. 
If $\deltaalpha < D_0$, the previous smoothed value is retained to avoid unnecessary re-modulations.
If $\deltaalpha \geq D_1$, the previous value is retained because rapid variations would otherwise lead to poorly estimated statistics.
When $\alphahat$ changes moderately, $\alphatilde$ is updated accordingly.

At every frame, the current estimates of the spectral-spatial covariance matrices are updated with the new data.
For example, $\hat{\v{S}}_{\v{x}}(\CyclicSet(\ell), \ell)$ is updated as:
\begin{align}
\scalemath{0.95}{
    \hat{\v{S}}_{\v{x}}(\CyclicSet(\ell), \ell) \leftarrow (1 \shortminus \beta) \hat{\v{S}}_{\v{x}}(\CyclicSet(\ell\shortminus1), \ell \shortminus 1) + \beta\,\v{x}(\ell) \v{x}(\ell)^H,
    }
\end{align}
where the value of the constant $\beta$ is given in \cref{ssec:experiments_real}.
Notice that the covariance matrices at different time frames may be functions of different modulation frequencies, thus the update is only approximately valid if the change in $\alphatilde$ is small and $C$ is the same.
In \cref{sec:experiments}, we see that if $\alphatilde$ changes slowly, as with the synthetic and instruments signals, the statistics are well estimated, and the cMWF performs well.
For real speech, $\alphatilde$ and $C$ vary over time, complicating statistics estimation.

\section{Experiments}\label{sec:experiments}
This section evaluates the proposed cMWF on simulated data, recordings from musical instruments, and speech signals.
The blind beamformer in \cref{eq:cmwf_blind}, which estimates the target covariance matrix through GEVD of $(\hat{\v{S}}_{\v{x}}, \hat{\v{S}}_{\v{v}})$, is compared against two unrealizable estimators that have access to ground-truth statistics. 
Results are given in terms SI-SDR improvements with respect to the unprocessed input at the first microphone \cite{roux_sdr_2019}.
The first oracle estimator, ``cMWF+", uses the ACP estimate $\hat{\v{S}}_{\v{d}}$ of the ground-truth target instead of its GEVD estimate. It is given by:
\begin{align}\label{eq:cmwf_oracle}
    \hat{\v{w}}_{\text{cMWF}}^{\text{+}} = (\hat{\v{S}}_{\v{d}} + \hat{\v{S}}_{\v{v}} + \lambda \v{I})^{-1} \hat{\v{S}}_{\v{d}}\, \v{e}_0.
\end{align}
The second unrealizable estimator, termed ``cMWF++", has access to the ACP estimate of the cross-statistics between the noisy and the ground truth target signals. It is given by:
\begin{align}\label{eq:cmwf_super_oracle}
    \hat{\v{w}}_{\text{cMWF}}^{\text{++}} = \hat{\v{S}}_{\lambda}^{-1} \hat{\v{s}}_{\v{x}\fourier{s}}.
\end{align}
To obtain similar variants of the narrowband MWFs, the spectral-spatial covariance matrices in \cref{eq:cmwf_blind,eq:cmwf_oracle,eq:cmwf_super_oracle} are replaced by the corresponding SCMs.
The amount of diagonal loading is calculated as in \cref{eq:diag_loading}, with
$\lambda_{\text{min}} = \num{e-9}$ and $\lambda_{\text{max}} = \num{e-4}$.
When evaluating \cref{eq:diag_loading}, we replace $\hat{\v{R}}_{\v{d}}$ with $\hat{\v{S}}_{\v{d}}$ for the cMWF variants.
The cMWF variants are only used for frequency bins $\omega_k$ that lie close to the harmonic frequencies in $\CyclicSet$, \ie satisfying $|\omega_k - \alpha_{\cycidx}| < \varepsilon \deltaomega$ for some $\alpha_{\cycidx} \in \CyclicSet$, where $\deltaomega$ is the spectral resolution and $\varepsilon$ is chosen as $\varepsilon=1.5$.
The remaining bins are processed with the narrowband MWF. 
We simulate a target point source, a directional interferer emitting white Gaussian noise (WGN) at $\SI{-10}{\decibel}$ SNR measured at the reference microphone, and spatially uncorrelated WGN at $\SI{30}{\decibel}$ SNR.
The RIRs for the target and interferer are randomly selected from a set of 26 RIRs measured in a room with $\text{RT60}=\SI{0.61}{\second}$ and $\SI{8}{\centi\meter}$ microphone spacing from the Bar-Ilan dataset \cite{hadad_multichannel_2014}.
Unless otherwise stated, we use $M=2$ microphones, $C=5$ frequency shifts, $K=512$ points for the FFT, and the square-root Hann window with $75\%$ overlap.
The noise covariance matrix $\v{S}_{\v{v}}$ used in the GEVD is estimated from a separate, \SI{2}{\second} long realization of the noise. 
Results are averaged over 50 Monte Carlo runs for the synthetic data experiments and over 10 runs for the real data experiments. 
We use different RIRs, noise, and target realizations in each run.
The plot lines indicate the mean values, while shaded areas represent the 95\% confidence intervals. 

\subsection{Synthetic data}\label{ssec:experiments_synthetic}
First, we evaluate the accuracy of the beamformers on simulated data.
The target signal is generated according to a simplified harmonic model \cite[Eq.~\!(7)]{bologni_harmonics_2024}, where the components at the different frequencies are perfectly correlated up to a multiplicative constant and a phase term. 
It is given by
$
Y(n) = B(n) \sum_{h=1}^H D_h \cos{(\omega_0 n h + \phi_h)},
$
where $\proc{B(n)}$ is a WSS process that describes the amplitude fluctuations over time, and the $D_h$ and the $\phi_h$ are random variables representing the relative amplitude and the phase of the sinusoids.
The process $\proc{B(n)}$ comprises independent Gaussian random variables distributed as $\mathcal{N}(0.5, 10)$ and lowpass filtered by a 4th order Butterworth filter with cutoff frequency $f_c = \SI{5}{\hertz}$. 
The $\phi_h$ are drawn from a uniform distribution $\mathcal{U}(-\pi, \pi)$, the $D_h$ are drawn from $\mathcal{U}(1, 10)$, and the frequency $\omega_0$ is drawn from $2\pi \cdot \mathcal{U}(60, 250)$. 
The number of harmonics $H$ is chosen as large as possible but obeys $\omega_0 H < f_s/2$. 
For the synthetic data experiments, the statistics are estimated using the entire signals, and $\alpha_1$ is assumed to be known.
Each generated audio sample lasts \SI{5}{\second}.
\begin{figure}[t]
    \vspace{-0.7cm}
    \centering
    \subfloat{%
    \includegraphics[width=0.6\linewidth]{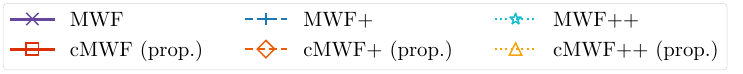}}
    \\
    \setcounter{subfigure}{0}
    \subfloat[ ]{%
    \includegraphics[width=0.49\linewidth]{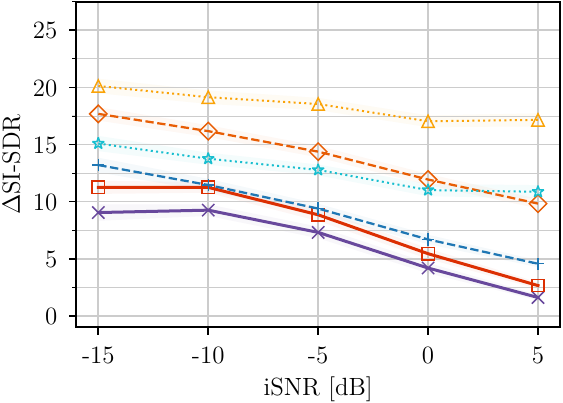}
    \label{fig:sim_snr}}
    \hfill%
    \subfloat[ ]{%
    \includegraphics[width=0.49\linewidth]{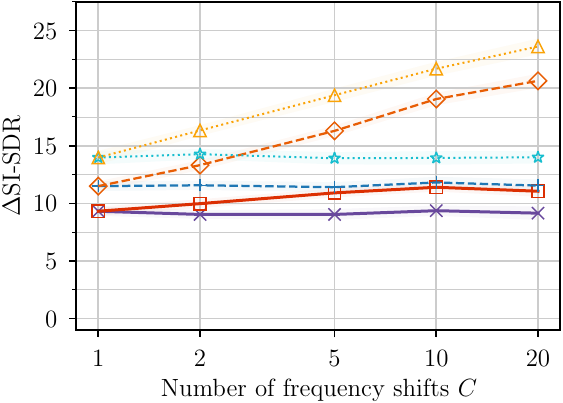}\label{fig:sim_C}}
    \\
    \subfloat[ ]{%
    \includegraphics[width=0.49\linewidth]{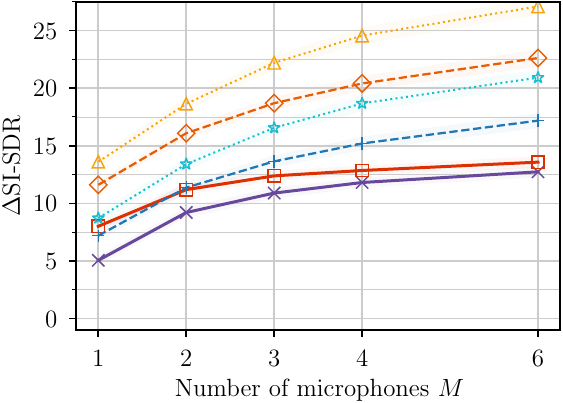}\label{fig:sim_mics}}
    \hfill%
    \subfloat[ ]{%
    \includegraphics[width=0.49\linewidth]{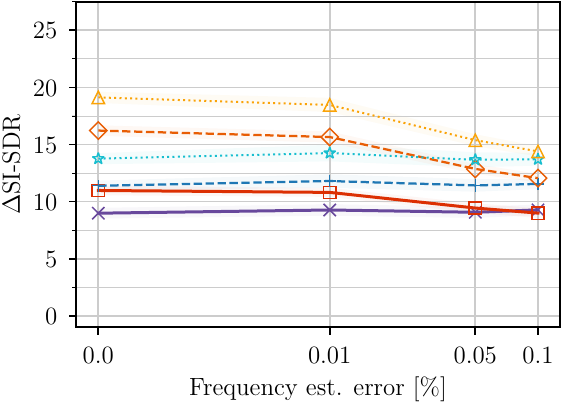}\label{fig:sim_f0}}
    \caption{Synthetic data. SI-SDR improvements over the noisy input for different beamformers. Each figure corresponds to a different varying parameter.}
    \label{fig:sim}
\end{figure}
In \cref{fig:sim_snr}, we vary the input SNR due to the interferer (iSNR). Results indicate that the cyclic beamformers always improve performance over the conventional MWFs.
\Cref{fig:sim_C} shows how the SI-SDR increases with the number of frequency shifts $C$ in the cyclic models.
Notice that, for $C=20$ frequency shifts, ``cMWF+" and ``cMWF++" are approximately \SI{10}{\decibel} SI-SDR better than ``MWF+" and ``MWF++", respectively.
As mentioned earlier, if only $C=1$ shift is considered, the cMWFs reduce to the corresponding MWFs.
Next, we vary the number of microphones $M$ in \Cref{fig:sim_mics}.
The performance of all beamformers improves when more microphones are available, as expected.
Finally, \cref{fig:sim_f0} analyzes the sensitivity of the cMWF to a bias applied to the fundamental frequency used to compute the cyclic set $\CyclicSet$ in \cref{eq:cyc_set_exp}.
The perturbed fundamental frequency is given by $\dot{\alpha}_1 = \alpha_1(1 + \alpha_{\text{err}}/100)$.
The cyclic beamformers are only beneficial if the error in the fundamental frequency is less than $0.1\%$.
Similarly, we found that the performance of the cMWFs degrades if the harmonics of the signal are not found at the exact integer multiples of the fundamental frequency (results not shown).

\subsection{Real data}\label{ssec:experiments_real}
\begin{figure}[t]
    \centering
    \vspace{-0.7cm}
    \subfloat{%
    \includegraphics[width=0.6\linewidth]{pics/legend.pdf}}
    \\
    \setcounter{subfigure}{0}
    \subfloat[]{%
    \includegraphics[width=0.49\linewidth]{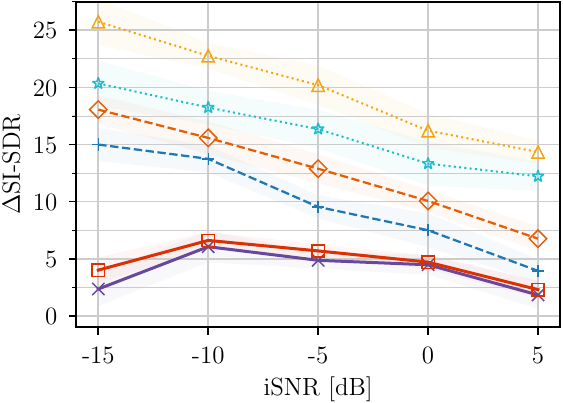}\label{fig:instr}}
    \hfill%
    \subfloat[]{%
    \includegraphics[width=0.49\linewidth]{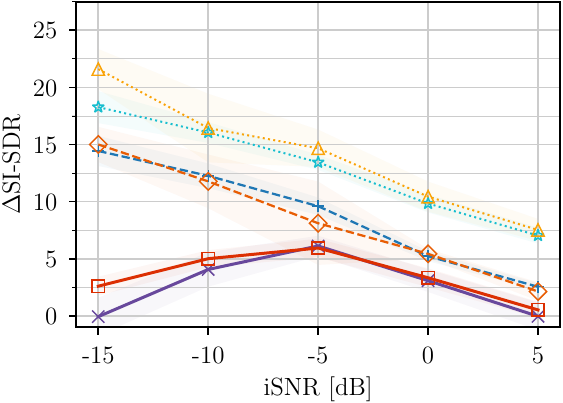}\label{fig:speech}}
    \caption{Real data. SI-SDR improvements over the noisy input for the different beamformers. (a) shows results on the IOWA dataset and (b) on speech data.}
    \label{fig:real}
\end{figure}

Next, we evaluate the recursive implementation of the algorithms described in \cref{ssec:rec_avg} using music or speech recordings as targets. 
The first dataset comprises single-note brass instrument samples from \cite{iowa}. When both vibrato and no-vibrato recordings are available, we select the latter ones. Only notes in the range C2 to C4 are considered, roughly corresponding to \SIrange{65}{260}{\hertz}.
The second dataset consists of real speech recordings from the TIMIT database, uttered by either a male or a female speaker.
In both experiments, we use $\beta=0.05$ for recursive averaging of covariance matrices.
In each Monte Carlo simulation, we randomly select $\SI{1}{\second}$ of data.
This value is chosen to be small because the single-note instrumental samples are of short duration.  
The constants that determine the update rate of $\alphatilde$ are set to $D_0=0.005$ and $D_1=0.2$. 
The fundamental frequency is estimated from the clean recordings.
The cMWF variants are employed only when the fundamental frequency $\alphahat$ has not changed significantly in the last frame to minimize the impact of poorly estimated covariance matrices; in other words, if $\deltaalpha \geq D_1$, we use the corresponding MWF variant for that time-frame.
Improvements in SI-SDR are measured for different input SNRs and shown in \cref{fig:real}.
The cMWF variants consistently outperform the benchmark for instrument recordings (\cref{fig:instr}), especially at lower iSNRs. 
For speech data (\cref{fig:speech}), the blind cMWF has a better SI-SDR score for lower iSNRs, and it performs similarly to the benchmark for iSNR \SI{-5}{\decibel} or higher.
PESQ \cite{rix_perceptual_2001} and STOI \cite{taal_algorithm_2011} results are omitted due to space limitations; they follow trends similar to SI-SDR.
The non-blind variants of cMWF perform erratically on speech data. By analyzing the output spectrograms (not shown), we hypothesize that this is due to extremely large output values that sometimes occur when the fundamental frequency changes.
In general, the reduced gains compared to \cref{ssec:experiments_synthetic} can be attributed to the limited accuracy in estimating $\alpha_1$ and the high variability of the fundamental frequency in speech. 
Additionally, whereas narrowband spatial covariance matrices change over time only by a scalar factor when sources are not moving, spectral covariance matrices vary with each note or phoneme, complicating their estimation.

\section{Conclusion}
This work proposed a cyclic multichannel Wiener filter for acoustic beamforming derived from a cyclostationary signal model.
The beamformer exploits spectral correlation across harmonic frequencies in the target signal to enhance performance.
While substantial SI-SDR gains are observed when the statistics, fundamental frequency, and harmonics are known exactly, as with synthetic data, the improvements are more modest on real recordings.

\bibliographystyle{IEEEtran}
\bibliography{refs25}







\end{document}